\documentclass[a4paper,11pt]{article}
\usepackage{pos}

\title{Effect of inelastic scattering on cosmic-ray-boosted dark matter}

\author[a]{Richie Diurba}
\author*[b,c]{Helena Kole\v{s}ov\'{a}}
\author[b]{Gailyn Monroe}

\affiliation[a]{LHEP, Bern University,\\
  Sidlerstrasse 5, CH-3012 Bern, Switzerland}

\affiliation[b]{Department of Mathematics and Physics, University of Stavanger,\\
4036 Stavanger, Norway}

\affiliation[c]{Supported by the Research Council of Norway under the grant no. 335388.\\
\,}

\emailAdd{richard.diurba@unibe.ch}
\emailAdd{helena.kolesova@uis.no}
\emailAdd{gailymonro@gmail.com}

\abstract{Sub-GeV dark matter particles evade standard direct detection studies since their typical energies in the galactic halo do not allow for detectable recoil of the heavy nuclei in the detectors. However, it was noted that if the dark matter particles have sizable couplings to nucleons, they can be boosted by interactions with galactic cosmic rays and probed by experiments like Xenon-1T. We revisit the resulting bounds on DM-nucleon interaction and improve on previous works by considering the DM-nucleus inelastic cross sections provided by the GENIE interaction event generator. Including inelastic scattering in the process of dark matter boosting increases the flux of high-energy dark matter coming to Earth. Additionally, allowing for inelastic scattering with nuclei is important for a realistic description of the dark matter interacting in the Earth's crust. We demonstrate these effects on a benchmark model where dark matter interactions are mediated by a dark U(1) vector boson.}

\FullConference{42nd International Conference on High Energy Physics (ICHEP2024)\\
18-24 July 2024\\
Prague, Czech Republic\\}

\begin{document}
\maketitle

\section{Introduction}


Novel constraints on the sub-GeV dark matter based on dark matter (DM) accelerated by galactic cosmic rays (CR) were derived in~\cite{Bringmann:2018cvk}, and variations of this idea were studied in a range of follow-up works. In this conference contribution, we 
pay particular attention to the effect of inelastic scattering of DM with nuclei. The latter becomes relevant in the context of cosmic-ray boosted dark matter (CRDM), where interactions with considerable momentum transfers ($Q^2$) are possible. In this regime, elastic coherent scattering of DM with nuclei gets suppressed by the nuclear form factors and with increasing $Q^2$, the DM particles first probe individual nucleons (``quasi-elastic scattering''), then excite nucleon resonances and, eventually, probe individual partons via deep inelastic scattering (DIS). 
To properly model the complicated nuclear effects, we employ the GENIE interaction event generator~\cite{Andreopoulos2009rq,Berger:2018urf}.


We concentrate on the effect of inelastic scattering on DM boosting by CR 
and on the attenuation of the CRDM flux in the Earth's crust, and compare the resulting exclusion limits with the previous analysis of~\cite{Alvey:2022pad}. Our study still assumes CRDM detection from coherent scattering, inclusion of inelastic scattering in the detection process is left for future work. 






\section{Results}
\label{sec:res}
Our ambition is to describe the inelastic scattering of DM with nuclei, including the DIS process. Consequently, we need to choose a concrete model for DM-quark interaction. We demonstrate our point using a benchmark model with a vector mediator $Z'$ considered previously in~\cite{Alvey:2022pad} and implemented in GENIE~\cite{Berger:2018urf}. We choose $m_{Z'}=1\,$GeV and fix the dark U(1) charges to be equal for all relevant quarks ($u$, $d$, $s$, $c$). We quantify the interaction strength by the DM-nucleon cross section in the highly non-relativistic limit $\sigma_{\rm SI}^\mathrm{NR}$, see~\cite{Alvey:2022pad} for more details. 

\begin{figure}
   \centering
    \includegraphics[width=.5\linewidth]{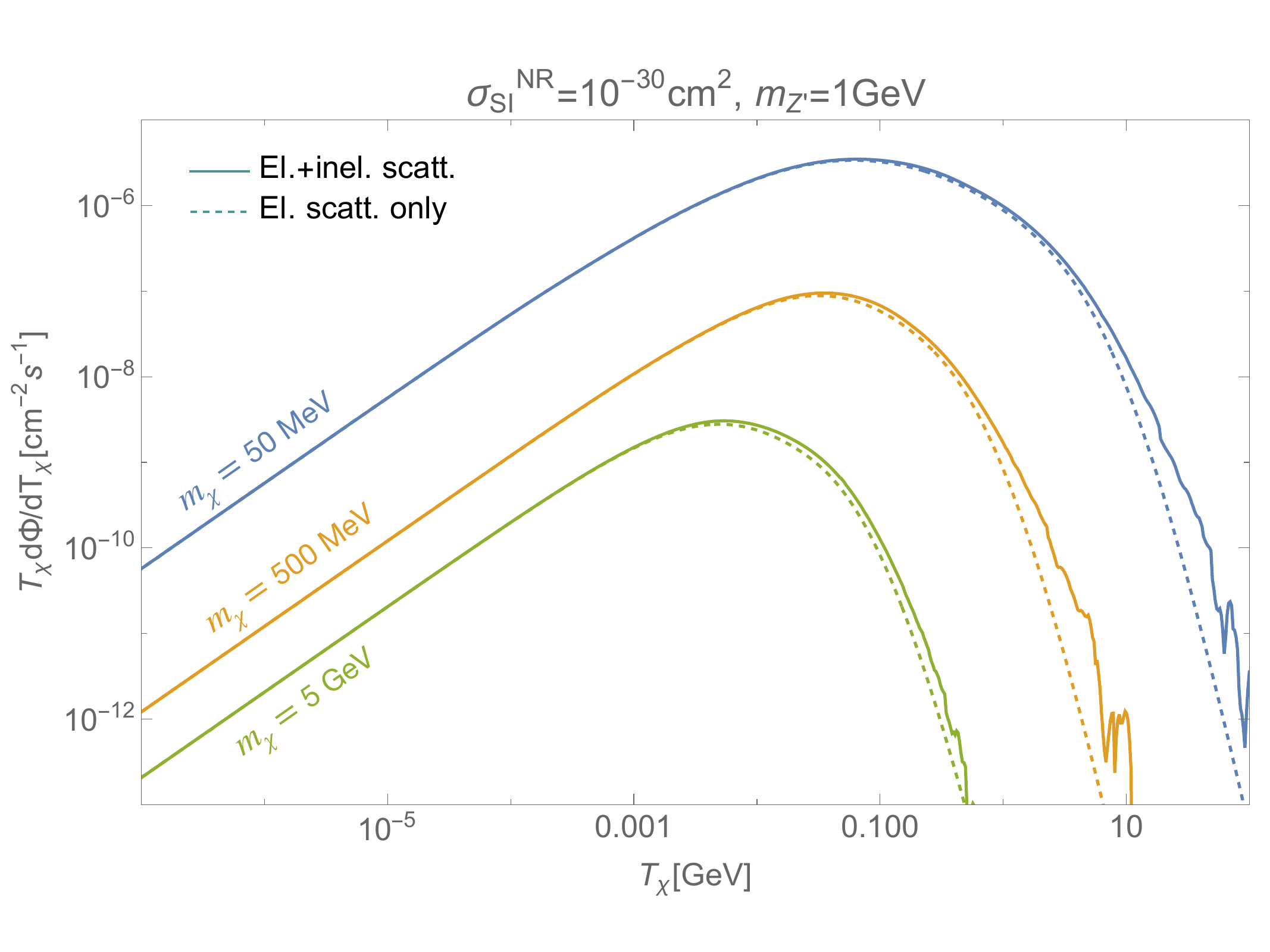}\includegraphics[width=.5\linewidth]{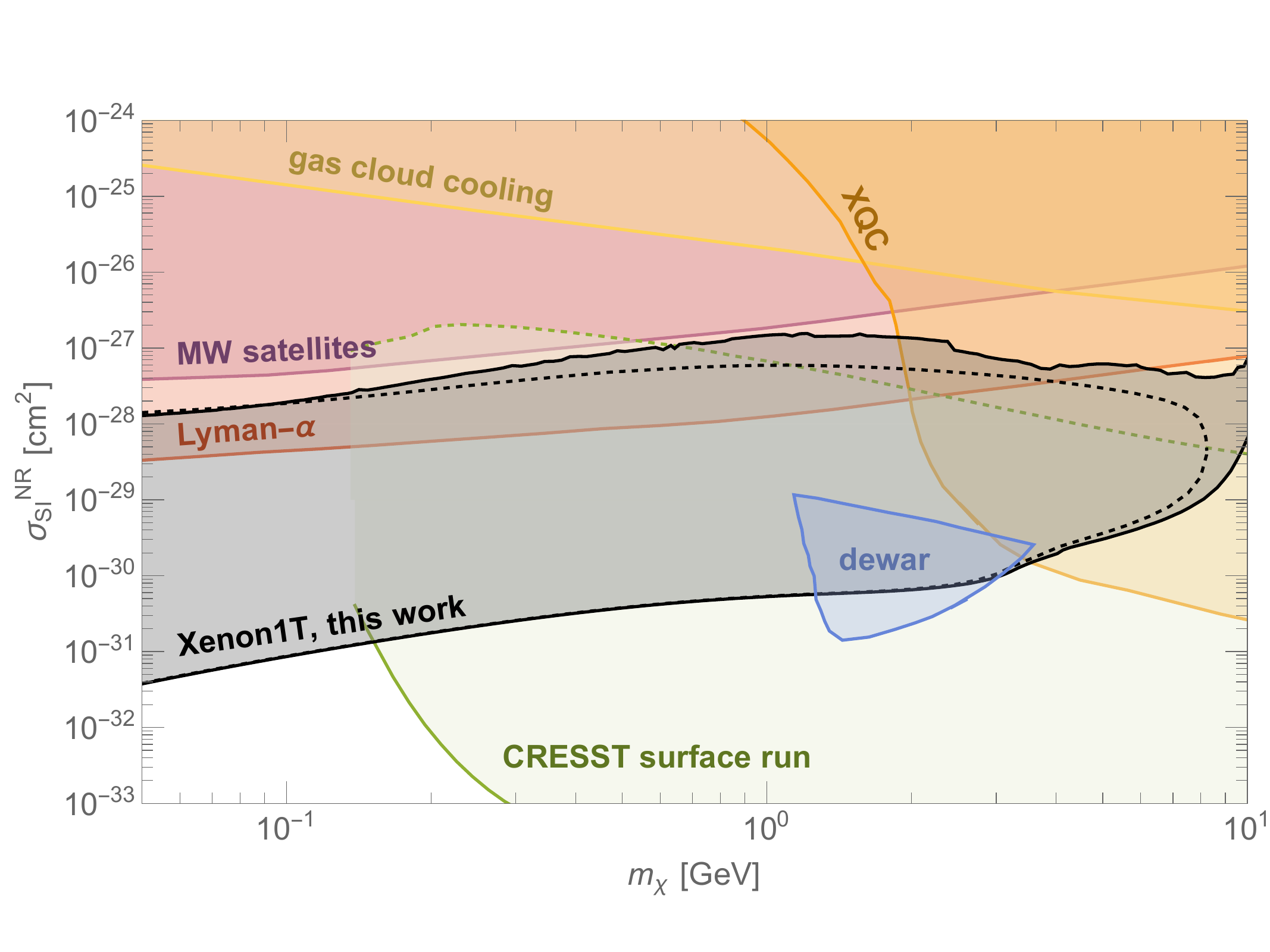}
    \caption{\emph{Left: CRDM flux at the Earth's atmosphere.} The solid and dashed lines depict the cases with and without the CR-DM inelastic scattering in the boosting process, respectively. \emph{Right: excluded parameter space.} The black-shaded region shows the region excluded by Xenon-1T experiments within our refined treatment, while the dashed line was obtained previously in ~\cite{Alvey:2022pad}. 
    See ~\cite{Alvey:2022pad} also for explanations of the complementary probes. Both plots were obtained using an extension of the DarkSUSY code~\cite{Bringmann:2018lay}.}
    \label{fig:res}
\end{figure}

In the left panel of Fig.~\ref{fig:res}, we present the CRDM flux for different DM masses and $\sigma_{\rm SI}^\mathrm{NR} = 10^{-30}\,$cm$^2$. We observe that the flux increases when inelastic interactions are included. Notice that the increase in the peak region is mainly due to the inclusion of quasi-elastic scattering for DM interactions with CR helium. In contrast, the increase at the highest $T_\chi$ is largely due to DIS with CR protons.

In the right panel of Fig.~\ref{fig:res}, we show the region of parameter space excluded by non-observation of CRDM by Xenon-1T experiment~\cite{XENON:2020fgj}. The excluded region is bounded from above since the DM with very strong coupling to nucleons would be stopped in the Earth's atmosphere or crust and would not reach the underground detectors. Inelastic scattering is vital in this stopping process since the elastic scattering with nuclei is suppressed for high-energy DM due to nuclear form factors. In~\cite{Alvey:2022pad}, the inelastic cross sections were estimated based on neutrino-nucleus cross sections and only DM with kinetic energies $T_\chi<10\,$GeV was considered. In this work, we model the inelastic cross sections more reliably using the code suited for DM~\cite{Berger:2018urf} and include the effect of DM with $T_\chi$ up to 100\, GeV. Inelastic scattering of DM with CR leads to an extension of the excluded region to higher $\sigma_{\rm SI}^\mathrm{NR}$ for heavier DM. The inelastic scattering in the boosting process also leads to an extension of the excluded region to lower $\sigma_{\rm SI}^\mathrm{NR}$. The extension is negligible for light DM (e.g., increases from $3.75 \times 10^{-32}\,$cm$^2$ to $3.82 \times 10^{-32}\,$cm$^2$ for $m_\chi = 50$\,MeV), 
but more pronounced for heavier DM.

\section{Conclusions and outlook}
\label{sec:conc}

We showed that the inclusion of the inelastic scattering of DM with CR leads to a mild extension of the CRDM limits posed by direct detection experiments. These experiments are tuned to $\mathcal{O}$(keV) elastic nuclear recoils expected for halo DM; therefore, the high-energy tail of the CRDM flux is less important and the limits corroborate results from studies that used a simplified treatment~\cite{Alvey:2022pad}. For GeV-scale neutrino experiments, the inelastic scattering channels can be used for CRDM detection and high-$T_\chi$ DM becomes crucial, motivating future work with inelastic interactions. The resulting CRDM bounds for neutrino detectors will be explored in a separate publication.


\bibliographystyle{JHEP}
\bibliography{biblio.bib}



\end{document}